\documentclass[aps,prb,twocolumn,superscriptaddress,square]{revtex4-1} 
\setcitestyle{numbers,square}
\usepackage[T1]{fontenc} 
\usepackage[utf8]{inputenc}
\usepackage{amssymb}
\usepackage[flushleft]{threeparttable}
\usepackage{url} 

\usepackage{hyperref} 
\hypersetup{
    bookmarks=true,         
    unicode=true,          
    pdftoolbar=false,        
    pdfmenubar=true,        
    pdffitwindow=true,      
    pdftitle={My title},    
    pdfauthor={Author},     
    pdfsubject={Subject},   
    pdfnewwindow=true,      
    pdfkeywords={keywords}, 
    colorlinks=true,       
    linkcolor=red,          
    citecolor=black,        
    filecolor=magenta,      
    urlcolor=cyan           
}

\usepackage[all]{hypcap} 
\usepackage{graphicx} 
\usepackage{amsmath} 
\usepackage{natbib} 
\usepackage{layouts}
\usepackage{verbatim}

\usepackage{mathtools}

\begin{document} 

\title{Energetics of the AK13 Semi-Local Kohn-Sham Exchange Energy Functional}

\author{A. Lindmaa} 
\email{alexander.lindmaa@liu.se}
\affiliation{Department of Physics, Chemistry and Biology (IFM), Link\"oping University, SE-581 83  Link\"oping, Sweden}
\author{R. Armiento}
\email{rickard.armiento@liu.se}
\affiliation{Department of Physics, Chemistry and Biology (IFM), Link\"oping University, SE-581 83 Link\"oping, Sweden} 
\date{\today} 

\begin{abstract} 
The recent non-empirical semi-local exchange functional of Armiento and K\"ummel, the AK13 [PRL 111, 036402 (2013)] incorporates a number of features reproduced by higher-order theory. The AK13 potential behaves analogously with the discontinuous jump associated with the derivative discontinuity at integer particle numbers. Recent works have established that AK13 gives a qualitatively improved orbital description compared to other semi-local methods, and reproduces a band structure closer to higher-order theory. However, its energies and energetics are inaccurate. The present work further investigates the deficiency in energetics. In addition to AK13 results, we find that applying the local-density approximation (LDA) non-self-consistently on the converged AK13 density gives very reasonable energetics with equilibrium lattice constants and bulk moduli well described across 14 systems. We also confirm that the attractive orbital features of AK13 are retained even after full structural relaxation. Hence, the deficient energetics cannot be a result of the AK13 orbitals having adversely affected the quality of the electron density compared to that of usual semi-local functionals; an improved orbital description and good energetics are not in opposition. We also prove that the non-self-consistent scheme is equivalent to using a single external-potential dependent functional in an otherwise consistent KS-DFT scheme. Furthermore, our results also demonstrate that, while an internally consistent KS functional is presently missing, non-self-consistent LDA on AK13 orbitals works as a practical non-empirical computational scheme to predict geometries, bulk moduli, while retaining the band structure features of AK13 at the computational cost of semi-local DFT.
\end{abstract}  

\maketitle 
\section{Introduction} 

Density-functional theory (DFT) \cite{hohenberg_inhomogeneous_1964}, in its standard formulation of Kohn and Sham (KS) \cite{kohn_self-consistent_1965}, is a very successful approach to the many-electron problem. It is generally accurate, and requires relatively little computational effort, compared to, e.g., wave-function based methods. While it is exact in principle, its key ingredient, the exchange-correlation (xc) energy functional is in practical calculations approximated. Many viable approximations have been proposed. 

A very common class of functionals rely solely on semi-local information in the density, i.e., $\nabla^{k} n$ to some order $k$, where $n(\textbf{r})$ is the electronic density. Semi-local DFT gives good results for structural properties. 
It is well established that there are a number of highly relevant exchange features  that semi-local DFT omits, such as the derivative discontinuity (DD) \cite{PhysRevLett.49.1691} and the relative offset between potentials for well-separated subsystems \cite{PhysRevA.54.1957}. These are closely related to the self-interaction error and lead, e.g., to over-delocalized KS orbitals, and which arguably constitutes a major deficiency in semi-local approximations. The missing features are to a large degree restored in higher-order methods such as hybrids \cite{becke_1993}, exact exchange (EXX) with the optimized effective potential (OEP) \cite{PhysRev.90.317, PhysRevA.14.36}, and many-body perturbation theory (in particular, the GW approximation \cite{PhysRev.139.A796}). These methods, however, increase the computational cost. This is in many situations undesirable, and hence it is a worthwhile goal of broad interest for applications across a large number of scientific fields to pursue a way to incorporate the exchange features into semi-local DFT. 

Recently, one of us proposed a semi-local functional \cite{PhysRevLett.111.036402} (AK13) which achieves an orbital description more similar to that of higher order methods. It has been shown to improve the electronic structure, optical dielectric constants \cite{PhysRevB.91.035107}, and potential shape \cite{ct500550s, PhysRevB.91.165121}, compared to other semi-local functionals. However, its energetics and total exchange energies are generally worse \cite{PhysRevLett.111.036402, ct500550s}. In particular, bulk Al gives an equilibrium lattice constant $6\%$ larger than the exchange-only version of the functional by Perdew, Burke, and Ernzerhof \cite{Phys.Rev.Lett.77.3865} (PBE). In the present work we further investigate this deficiency in energetics. We find that the average error is about $12\%$ compared to the couple of per cent error for typical semi-local functionals. 

The AK13 functional differs significantly from usual functionals on generalized gradient form (GGA) in that its refinement function is strongly divergent with the reduced density gradient, which heavily influences the resulting orbitals.
With this background, one may ask if the deficiency in energetics stems from a misfeature directly in the AK13 electron density caused by qualitative difference in the orbitals, or, if the deficiency instead primarily stems from the energies assigned by the AK13 functional to those densities. This is a relevant question, because the former option means the deficiency would be inherently associated with the main beneficial property of the AK13 functional, making it very challenging to address. A main point of the present work is to show that this is not the case.

It is not clear how one can deduce the quality of the density for energetics from looking at the density itself.
However, it has been a standard practice in the field to test new functionals by investigating their action on a known good electron density, e.g., a density from the local density approximation (LDA). Here, we reverse that test, and compare the results of inserting self-consistent densities from different functionals into the LDA functional. This method shows if reasonable energetics are retained for the AK13 density. As one may expect, we see that AK13 densities in the LDA functional does not visibly alter energies on the scale of typical total energies. However, the same turns out not to be true for the energetics (though reasonable energetics are acheived).

Hence, the primary contributions in this work to the field of semi-local methods with improved exchange features are: (\emph{i}) we investigate more closely the deficiency in energetics of the AK13 functional, (\emph{ii}) we answer in the affirmative that the qualitative changes to the AK13 orbitals have not adversely affected its density in a way that prevents reproducing reasonable energetics; (\emph{iii}) we suggest a direction for how to proceed to alter the energy functional of AK13 without significantly changing its orbitals or density; and (\emph{iv}) our results suggest that \emph{in lieu} of a single consistent KS-DFT functional, post-corrected AK13 works in practice as a computational scheme that is predictive for crystal geometry while at the same time retains the band structure features of AK13 with, e.g., increased KS band gaps more similar to those of higher-order theories. We are only aware of prior works demonstrating the success of AK13 band structure for systems using experimental geometries. 
 
The rest of the paper is organized as follows: Section~II summarizes the derivation and properties of the AK13 exchange functional. In Sec.~III we show how the scheme of using LDA on converged AK13 orbitals is equivalent to a single consistent external potential-dependent xc functional. In Sec.~IV, we show the numerical results for atoms and solids. In Sec.~IV we discuss our results and their implications for a future KS DFT functional. We summarize and conclude our work in Sec.~VI.  
\section{The AK13 exchange functional} 
We begin by briefly revisiting the derivation and motivation of the AK13 functional \cite{PhysRevLett.111.036402}. In the Perdew, Parr, Levy and Balduz ensamble extension of DFT \cite{PhysRevLett.49.1691}, the total energy of a system with an electron density $n(\textbf{r})$ of $N = N_0 + \eta$ electrons, with $N_0 \in \mathbb{N}$ and $\eta \in [0,1)$ moving in some external potential $v(\mathbf{r})$ is 
\begin{equation}\label{eq:engy_func} 
E_v[n] = T_{\mathrm{s}}[n] + E_{\mathrm{H}}[n] + \int v(\mathbf{r})n(\mathbf{r})\, d^{3}r + E_{\mathrm{xc}}[n], 
\end{equation} 
where $T_{\mathrm{s}}[n]$ is the non-interacting kinetic energy, $E_{\mathrm{H}}[n]$ the classical Coulomb-interaction (Hartree) energy, and $E_{\mathrm{xc}}[n]$ the exchange-correlation energy. The last term includes all remaining quantum many-body effects that we seek to approximate. A minimization of Eq.~(\ref{eq:engy_func}) results in the KS equations, given by (Hartree atomic units are used throughout)
\begin{equation}
\left\{-\frac{1}{2} \nabla^2 + v_{\mathrm{H}}(\mathbf{r}) + v(\mathbf{r}) + v_{\mathrm{xc}}(\mathbf{r})\right \} \phi_i(\mathbf{r}) =\varepsilon_i \, \phi_i(\mathbf{r}),  
\end{equation} 
where $\phi_i(\mathbf{r})$ are the KS orbitals, each corresponding to the eigenvalue $\varepsilon_i$, and $v(\mathbf{r})$ the external potential from the atomic cores, $v_{\mathrm{H}}(\mathbf{r})$ the Hartree potential, and $v_{\mathrm{xc}}(\mathbf{r})$ the xc potential, which is defined as $\delta E_{\mathrm{xc}}[n]/\delta n(\mathbf{r})$, and which usually is separated into a sum of its parts, the exchange (x) and correlation (c).  

The starting point of the derivation of the AK13 functional is the observation that the model potential of Becke and Johnson \cite{becke_johnson_2006} (BJ), $v_{\mathrm{x}}^{\mathrm{BJ}}(\mathbf{r})$, reproduces a behaviour that stems from the derivative discontinuity, specifically it undergoes a constant but sudden shift when the particle number changes across integer particle numbers \cite{PhysRevB.77.165106}. This is due to that, (i) the expression given by BJ takes an asymptotic value far outside a finite system that depends on the eigenvalue of highest occupied KS orbital $\varepsilon_I$, where $i=I$ (HOMO), and, (ii) BJ have stated that the potential \emph{may be manually shifted} with the system-dependent constant $\chi = \lim_{|\mathbf{r}| \to \infty} v_{\mathrm{x}}^{\mathrm{BJ}}(\mathbf{r})$ to achieve an asymptotic value of zero. Various modifications of the BJ potential have been shown to give band structures closer to higher-order theory \cite{tran_blaha_schwarz_2007, PhysRevLett.102.226401}, polarizabilities \cite{PhysRevB.77.165106}, as well as atomic and molecular properties \cite{PhysRevB.81.115108,ct100448x}. In contrast, the main feature of the AK13 functional is a potential that reproduces the same asymptotic behavior, but from a consistent energy-potential pair that avoids any theoretical or practical issue with model potentials \cite{PhysRevA.88.052519}. 

A semi-local exchange-functional on generalized-gradient approximation (GGA) \cite{PhysRevB.33.8800} form is defined as 
\begin{equation} 
  E^{\mathrm{sl}}_{\mathrm{x}}[n]=A_{\mathrm{x}} \int n^{4/3}(\textbf{r})F(s)\, \mathrm{d}^3r
\end{equation}
so that $A_{\mathrm{x}}=-(3/4)(3/ \pi)^{1/3}$, and $F(s)$ is a function of the reduced density gradient 
\begin{equation}\label{eq:defs}
s=|\nabla n|/(2k_{\mathrm{F}}n),
\end{equation}
where $k_{\mathrm{F}}=(3\pi^2n)^{1/3}$ is the Fermi wave vector. The choice $F(s) \equiv 1$ corresponds to LDA exchange. 
In Ref.~[\onlinecite{PhysRevLett.111.036402}], the sought asymptotic behavior is shown to be fulfilled by
\begin{equation}{\label{eq:enh_fac}} 
F^{\mathrm{AK13}}(s) = 1 + B_1s\ln(1+s) + B_2s\ln[1+\ln(1+s)], 
\end{equation} 
where $B_1=(3/5)\mu_{\mathrm{GE}}+(8/15)\pi$, $B_2 = \mu_{\mathrm{GE}} - B_1,$ and $\mu_{\mathrm{GE}}=10/81$. This can be verified by inserting the asymptotic density outside a finite system \cite{PhysRevA.49.2421}, i.e., 
\begin{equation}\label{eq:asym_dens}
n(\textbf{r}) \to m_ICe^{-2\sqrt{-2\varepsilon_I}r}, 
\end{equation}  
where $C$ is a system-dependent constant, and $m_I$ is the occupation number of the HOMO orbital. The corresponding exchange potential 
\begin{equation}\label{eq:pot_asym} 
v^{\mathrm{AK13}}_{\mathrm{x}}(\textbf{r}) \to -A_{\mathrm{x}}\frac{1}{3}B_1n^{1/3}s \sim \sqrt{-\epsilon_I},
\end{equation} 
which thus shifts discontinuously as a new orbital is occupied. Similar to the BJ model potential, one could propose a shift of the potential to zero alignment by adding a constant $\chi$, i.e., 
\begin{equation}\label{eq:zero_align} 
   v^0_{\mathrm{x}}(\textbf{r}) = v^{\mathrm{AK13}}_{\mathrm{x}}(\textbf{r}) + \chi, 
\end{equation} 
which is defined as 
\begin{equation}\label{eq:ak13chi}
\chi = -\lim_{|\textbf{r}| \to \infty} v^{\mathrm{AK13}}_{\mathrm{x}}(\textbf{r}). 
\end{equation} 
It follows that the expected \cite{PhysRevLett.49.1691} asymptotic limit of $v^0_{\mathrm{x}}(\textbf{r}) \to 0$, as $r \to \infty$ is retrieved. An exact expression for $\chi$ in a finite system can be derived \cite{PhysRevLett.111.036402}, and is given by  
\begin{equation}   
\chi = \frac{A_{\mathrm{x}}^2 Q_{\mathrm{x}}^2}{2} \left(1 \pm \sqrt{1-\frac{4\varepsilon_I^{\mathrm{SL}}}{A_{\mathrm{x}}^2Q_{\mathrm{x}}^2}} \right), 
\end{equation} 
where the plus sign applies for eigenvalues $\varepsilon_I < 0$ and $Q_{\mathrm{x}} = (\sqrt{2}/(3(3\pi^2)^{1/3}))B_1$. However, as it was argued in Ref.~[\onlinecite{PhysRevLett.111.036402}] there is no \emph{formal} requirement in KS DFT to perform this shift. Without the shift, the energy functional and potential pair remains consistent. Leaving out the shift may seem sufficient, since it has no effect on the density, and only appears as a constant shift of equal magnitude of the eigenvalues. Nevertheless, \emph{if} a functional could be constructed to give a shifted potential, that would most likely affect the energies. Similarly, in this work, we explore the consequences of a scheme that alters the energy without changing the orbitals. This idea is somewhat formalized in terms of a single density and potential-dependent functional in the next section.  

\section{Non-self-consistent LDA energies from self-consistent AK13 orbitals} 
As we outlined in Secs.~I and~II, the aim of the present work is to investigate the energetics that result from a non-self-consistent evaluation of the LDA energy on orbitals obtained self-consistently with the AK13 functional. In practice, the implementation of this scheme is completely straightforward as an extra energy evaluation after electronic convergence in a regular AK13 calculation. In this section we discuss how this scheme can alternatively be seen as a single self-consistent functional, but which requires an external potential dependence.

Our starting point is a general, non-unique partition of the xc-functional 
\begin{equation}\label{eq:division}
E_{\mathrm{xc}}[n] = E^{\mathrm{sl}}_{\mathrm{xc}}[n] +  E^{\mathrm{nl}}_{\mathrm{xc}}[n], 
\end{equation}
consisting of semi-local (sl) and non-local (nl) parts. This relation defines the functional $E^{\mathrm{nl}}_{\mathrm{xc}}[n]$. It is allowed to be non-local in the sense that it may depend on the density $n$ throughout the entire system. Furthermore, we let $E^{\mathrm{sl}}_{\mathrm{xc}}[n] = E^{\mathrm{sl}}_{\mathrm{x}}[n] + E^{\mathrm{sl}}_{\mathrm{c}}[n]$, and, from now on, take $E^{\mathrm{sl}}_{\mathrm{x}}[n] = E^{\mathrm{AK13}}_{\mathrm{x}}[n]$, where $E_{\mathrm{c}}^{\mathrm{sl}}[n]$ is a suitable choice of a semi-local correlation-energy functional.

For a given external potential $v$, define the density that results from a self-consistent solution of the KS DFT problem of only the sl part of the problem as $n^{\mathrm{sl}}[v]$, which via Eq.~(\ref{eq:defs}) also gives $s^{\mathrm{sl}}[v]$. It is now possible to make a definition of the contribution from $E_{\mathrm{xc}}^{\mathrm{nl}}$ to give \emph{exactly} LDA energetics as
\begin{multline}\label{eq:constant} 
   E^{\mathrm{nl}}_{\mathrm{xc}}[n,v] = -A_{\mathrm{x}}\int n^{4/3}(\textbf{r}) \bigg[ B_1s^{\mathrm{sl}}\ln(1+s^{\mathrm{sl}}) + \\
    B_2s^{\mathrm{sl}}\ln(1+\ln(1+s^{\mathrm{sl}})) \bigg] \, d^3r,    
\end{multline} 
and $E_{\mathrm{c}}^{\mathrm{sl}}[n] = E_{\mathrm{c}}^{\mathrm{LDA}}[n]$ [\onlinecite{PhysRevB.45.13244}].
The definition in Eq.~(\ref{eq:constant}) is chosen to make all but the first term in Eq.~(\ref{eq:enh_fac}) cancel out in the energy functional, but will by constraction not affect the orbitals. This definition is \emph{not} a valid KS DFT functional due to its explicit dependence on the external potential from $s^{\mathrm{sl}}$. However, such a dependence in the exchange-correlation functional is frequently seen in other schemes, and the present scheme is thus on equal fundamental footing with them. More importantly, Eq.~(\ref{eq:division}) gives an explicit construction idea to explore for a future functional within KS DFT with the sought properties. 


\section{Results on Atomic energies and the energetics of solids} 

\begin{figure}[htb] 
	\centering 
 	\includegraphics[width=0.49\textwidth, angle=0]{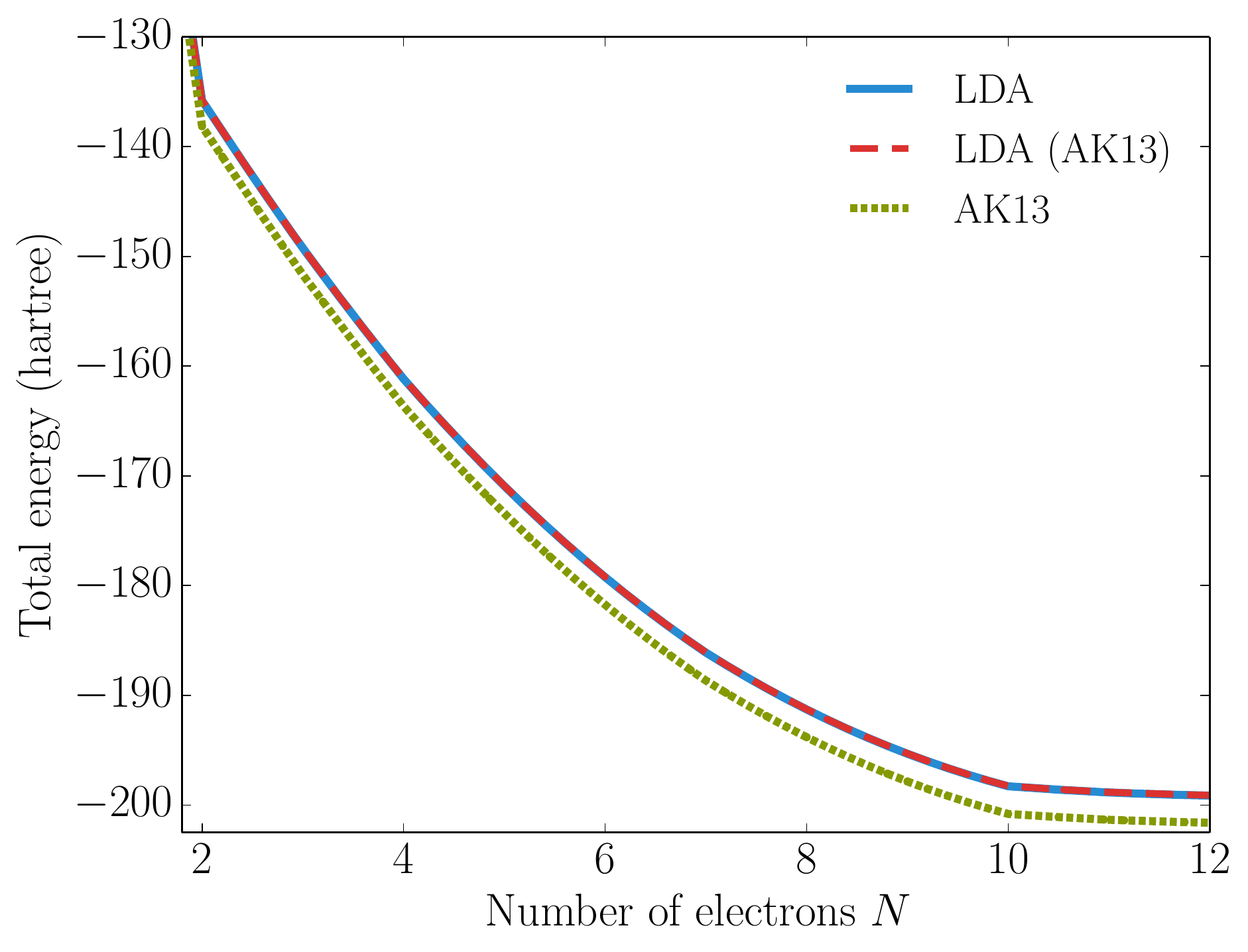}
  	\caption{Total energies versus electron occupation number for a Mg ion. In all cases, the total energy is obtained from fully self-consistent calculations using the PW LDA correlation-energy functional. Self-consistent LDA (solid blue) energies stay close to those calculated from LDA on converged AK13 orbitals (long-dashed red). Self-consistent AK13 (short-dashed green) consistently stays below the other energies.} 
 	\label{fig:fig1} 
\end{figure}

\begin{figure}[htb] 
	\centering 
 	\includegraphics[width=0.49\textwidth, angle=0]{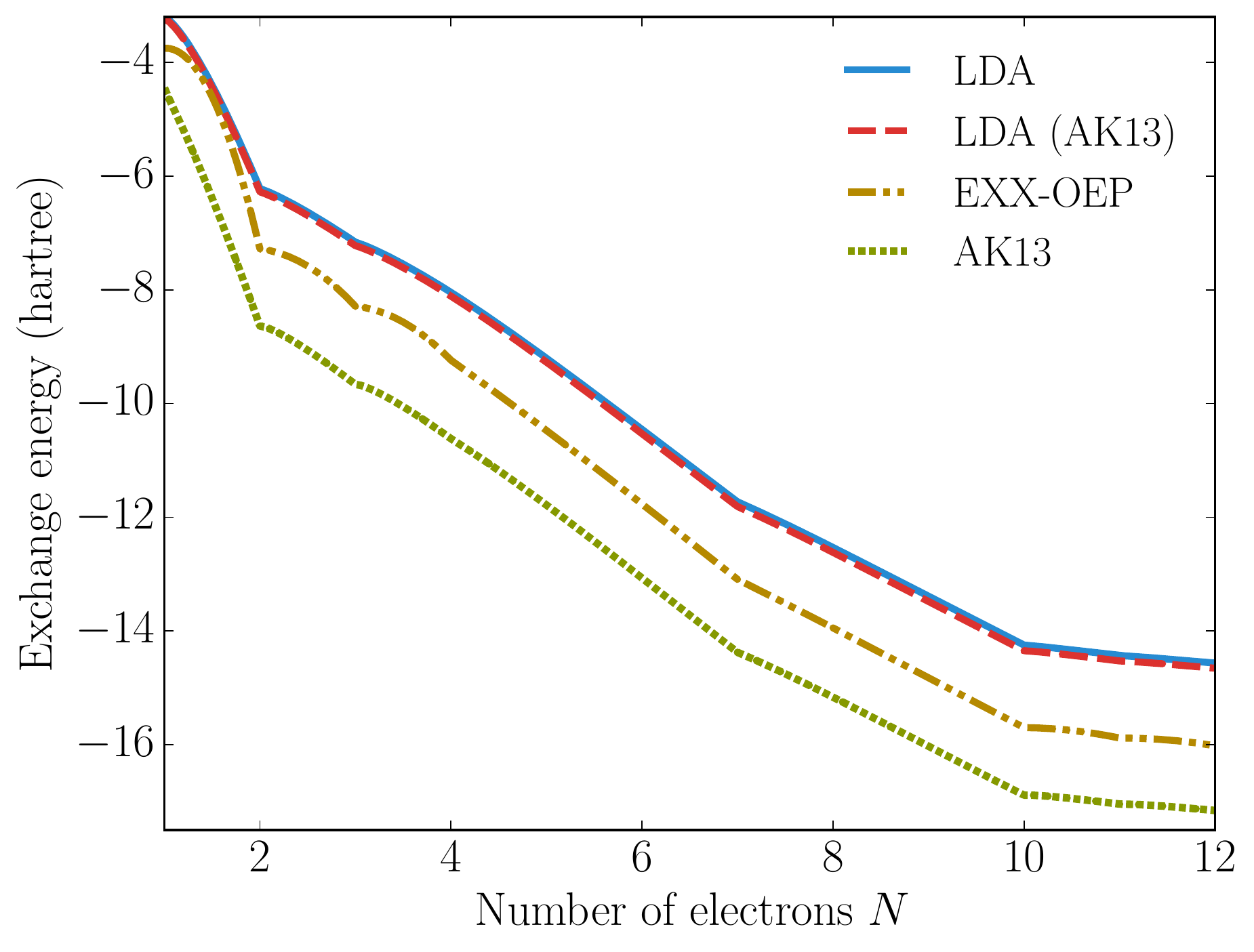} 
  	\caption{Total exchange energies versus electron occupation number for a Mg ion, as obtained from fully self-consistent exchange-only calculations. In similarity with Fig.~\ref{fig:fig1}, the self-consistent LDA energies (solid blue) are close to those calculated from LDA on converged AK13 orbitals (long-dashed red). The self-consistent AK13 energies (short-dashed green) are far below those of the ordinary, self-consistent LDA. For comparison, the result of exact-exchange OEP (yellow dashed-dotted) is also included.}
 	\label{fig:fig2}
\end{figure}

Our first numerical results shows the energy change by using non-self-consitent LDA on converged AK13 orbitals for total atomic energies. We use a modified atomic DFT code, originating from an early work of Talman and Shadwick \cite{PhysRevA.14.36}, to perform fully self-consistent spin-polarized DFT calculations on a one-dimensional grid of $800$ mesh points, $r \in \{ e^{-8+0.015i} \}^{799}_{i=0}$. Figure~\ref{fig:fig1} shows the total energy of a magnesium ion as a function of the electron occupation number, i.e, the system is successively filled with electrons, keeping the external potential of the atomic core fixed. In Figure~\ref{fig:fig2} we show the total exchange energies for the same system. Observe that the results depicted in Fig.~\ref{fig:fig2} have been obtained from exchange only calculations. 
In both cases, we find that self-consistent AK13 energies are consistently lower than those of both self-consistent LDA and exact exchange OEP (EXX-OEP). Furthermore, the energies calculated from LDA applied to the converged AK13 orbitals stay very close to those of the self-consistent LDA, i.e., for the LDA energy functional there is hardly any difference between using the LDA or AK13 orbitals. 

\begin{figure}[htb] 
	\centering 
 	\includegraphics[width=0.49\textwidth, angle=0]{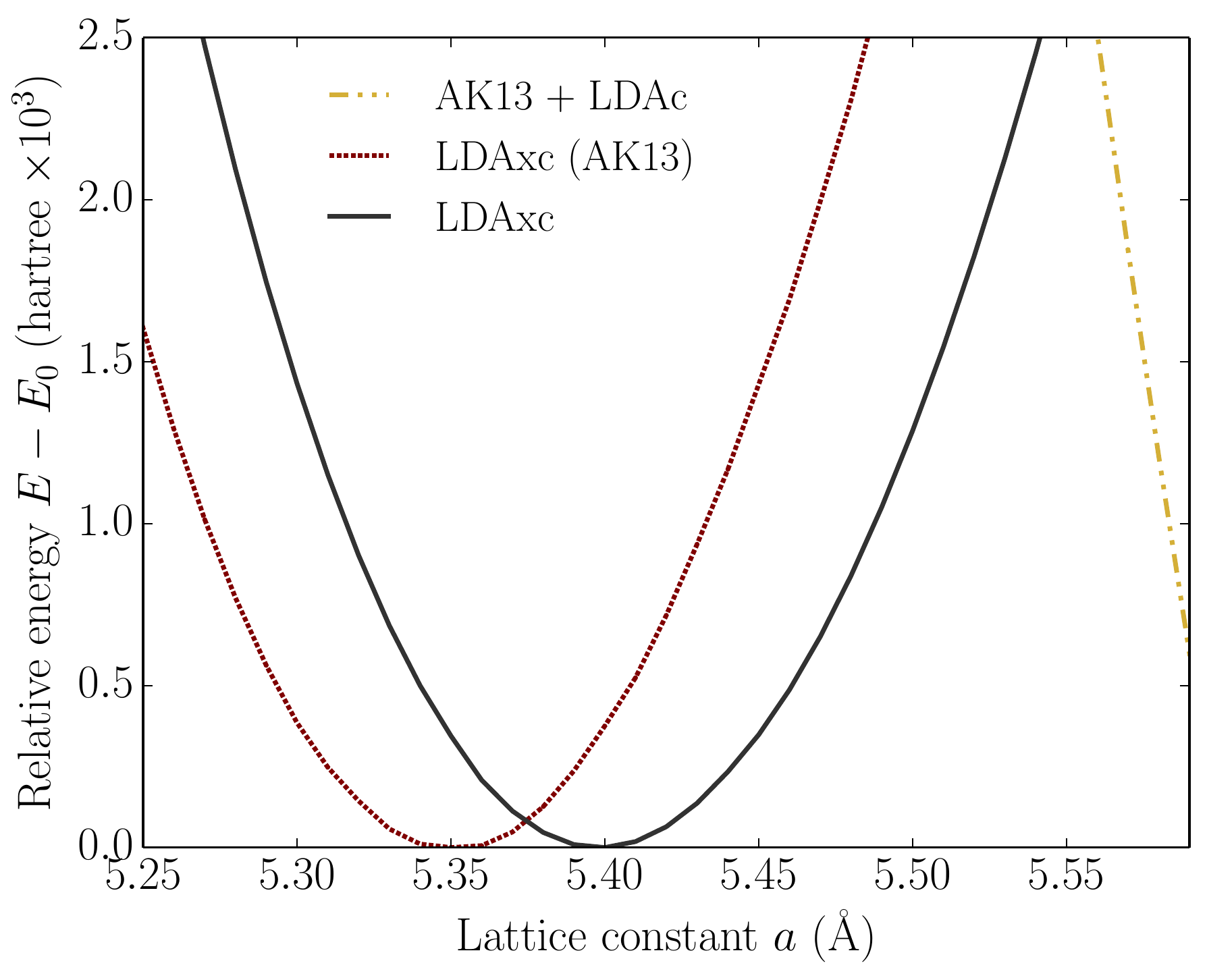} 
        \caption{Equation of state for silicon in the diamond structure. In all cases, we used the PW LDA correlation-energy functional. The use of self-consistent AK13 (yellow dashed-dotted) results in no minimum. By contrast, when using LDA on converged AK13 orbitals, an optimal lattice constant is found. The result of self-consistent LDA (solid black) The optimal lattice constant of LDA on converged AK13 orbitals (short-dashed red) is ca 0.05 Å less than that of self-consistent LDA (solid black).} 
 	\label{fig:fig3} 
\end{figure} 

\begin{figure}[htb] 
	\centering 
 	\includegraphics[width=0.49\textwidth, angle=0]{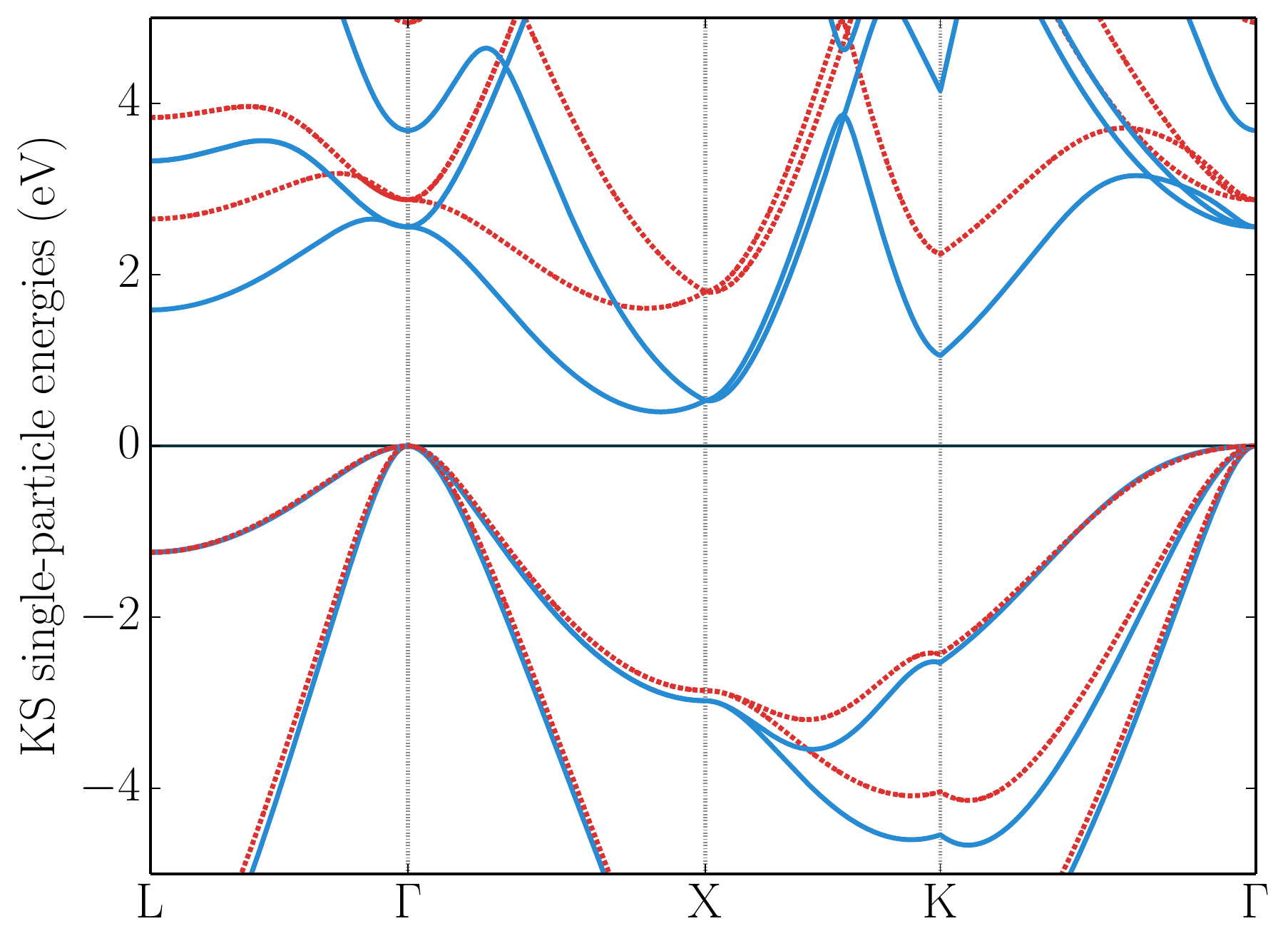} 
        \caption{Kohn-Sham band structure along high symmetry directions for silicon in the diamond structure, using self-consistent LDA (solid blue), and self-consistent AK13 (red dashed). The band structure has been calculated at equilibrium lattice constants. The appearance of the bands differs marginally from the ones calculated at the experimental lattice constant (not shown).} 
 	\label{fig:fig4} 
\end{figure} 

Next we consider periodic solids. To this end we use Elk, a full-potential, linearized augmented plane wave (LAPW) \cite{singh} code. For each solid, we calculate the total energy at 7 different volumes in an approximate interval of $\pm 10 \, \% $ around the equilibrium volume $V_0$. The obtained values we fit to the Birch-Murnaghan equation of state (EOS) \cite{PhysRevB.28.5480}, from which we obtain the optimal lattice geometry and bulk modulus. Specifically, the lattice parameters of this work were calculated by choosing volume points in the vicinity of the equilibrium volume as obtained from self-consistent LDA. Since the optimal volume points for AK13 would be at a range of larger volumes than the points optimal for LDA, our calculated AK13 equilibrium lattice parameters and bulk moduli are not as accurate as the values reported for the other methods. However, the accuracy is still sufficient to illustrate the poor performance of AK13 for these quantities, and we thus find it unnecessary to optimize the volume points for greater accuracy. In all calculations we take $E^{\mathrm{sl}}_{\mathrm{c}}[n]$ to be the parameterized LDA functional of Perdew and Wang (PW) \cite{PhysRevB.45.13244}. In all calculations we use $12\times12\times12$ $\mathbf{k}$-points, 50 empty bands, and a cutoff of $k_{\mathrm{max}} = 8.0 / R$, where $R$ is the muffin-tin radius. In Elk, the values of $R$ are set as default, which means that they are automatically rescaled when overlapping (however, the results were carefully checked so as to not be sensitive to minor changes in $R$). 

Results for silicon in the diamond structure are shown in Figs.~\ref{fig:fig3}, and~\ref{fig:fig4}. Figure~\ref{fig:fig3} shows how AK13 gives a much too large lattice minimum. When we use the orbitals of AK13 to calculate the non-self consistent LDA contribution, however, the result is a minimum much closer to self-consistent LDA. The shape of the energy vs.\ lattice constant curve in this case closely resembles that of self-consistent LDA. Furthermore, Fig.~\ref{fig:fig4} shows two different sets of KS bands, each of which has been calculated at the equilibrium lattice constant when using non-self-consistent LDAxc on AK13 orbitals. This, together with the result shown in Fig.~\ref{fig:fig3}, illustrates that the main features of the AK13 orbitals are unaffected by structural relaxation.  

To make sure Si is not a single fortuitous case, we repeat this analysis for 14 different solids and tabulate the equilibrium lattice constants $a_0$ and bulk moduli $B_0=-V\,\partial^2E/ \partial V^2|_{V=V0}$. 
The results are shown in Tables~\ref{table:table1} and \ref{table:table2}. For the lattice constants $a_0$, we find that non-self consistent LDA on AK13 orbitals gives lattice constants that are slightly smaller, with a mean error (me) of about -0.2 Å, than those obtained from self-consistent LDA. The bulk moduli have also been substantially improved compared to self-consistent AK13. The calculated values of $B_0$ are reasonably close to those of self-consistent LDA. In Table~\ref{table:table2}, we have also included the result from the Perdew, Burke, and Ernzerhof (PBE) functional for comparison. 

At the optimized geometries, we have calculated the KS gaps, $\Delta_{\mathrm{KS}}$. These are shown in Table~\ref{table:table3}.  
We emphasize that one cannot make a direct quantitative comparison between KS band gaps and experimental gaps. However, as has been argued previously \cite{PhysRevB.91.035107}, the significance of these results is that the present method retains the AK13 band structure that is more similar to that of higher-order theory. 

\section{Discussion} 
The results in Table~\ref{table:table1} demonstrate that the use of non-self-consistent LDA energies on converged AK13 orbitals results in reasonable values of equilibrium lattice constants (mean average error (mae) of 0.1 Å, mean average relative error (mare) of 0.02), although with a tendency to over-bind even somewhat more than self-consistent LDA. The over-binding seems to be a general trend among the studied systems, although somewhat surprisingly the reverse is true for Al. For the bulk moduli in Table~\ref{table:table2}, our approach gives an mae of 9.7 GPa (mare 0.3), which is a small difference, i.e., of the same magnitude that one can expect from using various implementations found in different computational codes.

The orbital features (i.e., features in the KS band structure and band gaps) appear to be more or less unaffected by the change in lattice constant due to structural relaxation using the LDA energy on converged AK13 orbitals as compared to AK13 on experimental geometries. While the quality of the DFT orbital description is a difficult topic, we have discussed in previous works \cite{PhysRevB.91.035107} how we find band gaps and orbital features calculated with various methods to end up in one of two categories. Either (\emph{i}) the orbital features have the typical LDA and PBE characteristics, e.g., with band gaps much smaller than the experimental ones and with a very reduced step structure in the KS potential for atoms; or (\emph{ii}) the orbitals share features of higher-order methods (in particular, exact exchange DFT), i.e., they are more similar to actual quasiparticle states and, as a result, give expanded band gaps and optical properties more similar to experiments, etc. The orbitals of AK13 are in the latter category, both when calculated on experimental geometries and, as shown in this work, on relaxed geometries (when using LDA for the energies.)

The stated main focus of the present work is to establish if the inaccurate energetics of the AK13 functional is an inherent feature of the changed orbitals, as compared to other, general semi-local functional approximations. In light of our findings above, this appears to \emph{not} be the case, since the LDA energy functional gives a physically acceptable energy minimum when applied to those orbitals. Furthermore, we have shown that the scheme of using AK13 to converge the orbitals and then LDA to calculate the energies is, in fact, equivalent to using a single consistent exchange-correlation functional, but, with an explicit external potential dependence, Eq.~(\ref{eq:constant}). From these results, we surmise that it is likely possible to find a density functional within KS DFT that is capable of incorporating both these features at the computational cost of semi-local DFT. We suggest that the external-potential dependent expression in Eq.~(\ref{eq:constant}) is a possible starting point for such a construct.

It is pertinent at this point to ask why the LDA energetics are slightly worsened when using orbitals we expect to be \emph{more accurate} than the LDA orbitals (see Fig.~\ref{fig:fig3} and Table~\ref{table:table1}). In the following we will speculate on a few possible reasons: (\emph{i}) the use of non-self-consistent LDA on converged AK13 orbitals is meant as an approximation at what a fully self-consistent KS DFT scheme can achieve, and thus the lack of a fully self-consistent scheme may itself lead to worsened energetics; (\emph{ii}) The performance of LDAxc relies on a cancellation of error between exchange and correlation \cite{kurth, dft_primer}. We have used LDA correlation when converging the AK13 orbitals, which is not ideal. A correlation functional better matched with AK13 exchange may influence the orbitals enough to recover LDA energetics or better; (\emph{iii}) The AK13 functional may yield an improved orbital description in the sense of reproducing exchange features missing from other semi-local functionals, but it was never optimized for accurate energies or energetics. It may be that the AK13 form with its non-empirically derived values for $B_1$ and $B_2$ in Eq.~(\ref{eq:enh_fac}) lead to other minor deficiencies in the orbitals that, in contrast to the main conclusion of this paper, inherently worsen energetics somewhat compared to LDA. However, if true, the conclusion would be that this effect is on a much smaller scale than what one is led to believe from the energetics of self-consistent AK13. 

\begin{table}\label{table:table1} 
\caption{Equilibrium lattice parameter $a_0$ (\AA) of 14 solids considered in this work. } 
\begin{ruledtabular}
\begin{threeparttable} 
\begin{tabular}{@{\ } l l l l l l @{\ }}
             & LDA    & PBE & AK13 & This work & Expt.\tnote{a}   \\
\hline 
Al           & $3.995$  &  $4.064$  & $4.286$ & $4.027$ &  $4.047$   \\           
Si          & $5.400$  &  $5.470$  &  $5.897$ & $5.351$  &  $5.430$   \\
$\alpha$-Sn   & $6.414$  &  $6.619$  & $7.021$ & $6.351$  &  $6.481$   \\
Ge       & $5.592$  &  $5.742$  & $6.521$ & $5.542$  &  $5.652$   \\
GaAs        & $5.611$  &  $5.759$  & $6.606$ &  $5.575$  &  $5.648$   \\
InAs          & $6.035$  &  $6.256$  & $6.985$ &$5.823$  &  $6.058$   \\ 
Ca            & $5.295$  &  $5.482$  & $6.144$ & $5.286$  &  $5.580$   \\
Li          & $3.363$  &  $3.492$  & $3.911$ & $3.131$ &  $3.477$   \\
K            & $5.046$  &  $5.285$  & $6.228$ & $4.727$  &  $5.225$   \\
LiCl         & $4.966$  &  $5.185$  & $6.357$ & $4.842$  &  $5.106$   \\
AlN         & $4.342$  &  $4.399$  & $4.649$ & $4.325$  &   $4.380$   \\
TiN          & $4.173$  &  $4.249$  & $4.453$ & $4.168$  &  $4.239$   \\
AlAs         & $5.633$  &  $5.727$  & $6.591$ &$5.578$  &  $5.661$   \\ 
Sr           & $5.781$ &  $6.003$  & $6.357$ & $5.700$  &  $6.080$    \\
\hline 
me              & $-0.101$  & $0.048$ & $0.639$ &  $-0.188$         &      \\
mae              & $0.101$  & $0.073$ & $0.639$  &  $0.188$         &      \\
mare             & $0.020$  & $0.013$ & $0.121$  &  $0.037$        &       \\ 
\end{tabular}
\begin{tablenotes} 
\item[a]Reference \onlinecite{ogilvie_1992,ogilvie_1993} 
\end{tablenotes}
\end{threeparttable} 
\end{ruledtabular} 
\end{table} 

\begin{table}\label{table:table2} 
\caption{Equilibrium bulk modulus $B_0$ (GPa) of 14 solids considered in this work.}
\begin{ruledtabular}
\begin{threeparttable}
\begin{tabular}{@{\ } l l l l l l@{\ }}
             & LDA    & PBE & AK13 & This work & Expt.\tnote{a}   \\
\hline 
Al           & $84.1$  &  $75.9$  & $68.9$ & $77.6$  &  $73$    \\           
Si           & $97.6$  &  $91.6$  & $51.7$ & $108$  &  $99.2$    \\
$\alpha$-Sn  & $42.2$  &  $33.1$  & $74.1$ & $44.7$  &  $53$    \\
Ge           & $68.8$  &  $55.9$  & $26.6$ & $75.0$  &  $75.8$    \\
GaAs         & $73.1$  &  $59.2$  & $23.2$ & $75.3$  &  $75.6$    \\
InAs         & $64.5$  &  $59.2$  & $29.2$ & $75.3$  &  $58$    \\
Ca         & $22.3$  &  $19.7$  & $15.8$ & $23.7$  &  $15$    \\ 
Li         & $15.0$  &  $7.58$  & $3.9$ & $17.0$  &  $13.0$    \\
K         & $4.42$  &  $3.56$  & $3.4$ & $9.49$  &  $3.7$    \\
LiCl         & $40.9$  &  $22.8$ & $11.2$ &  $49.1$  &  $35.4$    \\
AlN          & $211$  &  $197$  & $157$ & $218$  &  $202$    \\
TiN          & $324$  &  $278$  & $350 $ & $326$  &  $288$    \\
AlAs         & $73.7$  &  $64.1$  & $14.4$ & $78.6$  &  $82$    \\ 
Sr           & $15.8$ &  $12.3$  & $32.2$ & $18.6$  &  $12$    \\ 
\hline
me         & $3.7$  &  $-7.1$    & $-16.0$    & $7.9$           \\ 
mae         & $8.0$  &  $9.3$    & $30.9$    & $9.7$            \\ 
mare        & $0.17$     & $0.43$ & $4.11$    & $0.30$              \\ 
\end{tabular} 
\begin{tablenotes} 
\item[a]Reference \onlinecite{ogilvie_1992,ogilvie_1993} 
\end{tablenotes}
\end{threeparttable}
\end{ruledtabular}
\end{table} 

\begin{table}\label{table:table3}
\caption{Kohn-Sham gaps $\Delta_{\mathrm{KS}}$ (eV) for X solids calculated at equilibrium lattice constants $a_0$ from Table~\ref{table:table1}.} 
\begin{ruledtabular}
\begin{threeparttable}
\begin{tabular}{@{\ } l l l l l @{\ }}
             & LDA    & PBE & This work & Exp.   \\
\hline 
           Si           & $0.4$  &  $0.6$  &  $1.6$  &  $1.1$\tnote{c}    \\
$\alpha$-Sn  & $-0.1$  &  $-0.2$  &  $0.3$  &  $0.1$\tnote{c}    \\
Ge           & $0.1$  &  $0.0$  &  $1.2$  &  $0.7$\tnote{c}    \\
GaAs         & $0.4$  &  $0.1$  &  $1.8$  &  $1.4$\tnote{c}    \\
InAs         & $0.0$  &  $0.0$  &  $0.3$  &  $0.4$\tnote{d}    \\
AlAs         & $1.3$  &  $1.5$  &  $2.8$  &  $2.2$\tnote{d}    \\ 
\end{tabular}
\begin{tablenotes}
\item[c]Reference \onlinecite{PhysRevB.91.035107}  
\item[d]Reference \onlinecite{J.Appl.Phys.89.5815} 
\end{tablenotes}
\end{threeparttable}
\end{ruledtabular}
\end{table} 
  
\section{Summary and Conclusions}
In this work we have studied the deficiency of the energetics of the AK13 functional and, in doing so, taken a step towards a semi-local KS density functional that combines accurate energetics with the improved orbital description similar to that of higher-order theories. We have demonstrated that the description of exchange bonding in AK13 is generally highly unsatisfactory. We have shown that this deficiency cannot primarily be attributed to how the qualitatively different AK13 orbitals changes the electron density. Rather, one finds LDA-like energetics to be recovered when the self-consistent AK13 density is inserted into the LDA xc functional. We have also demonstrated that the changes in the AK13 band structure are robust under changes in geometry, and thus are retained when the structure undergoes structural relaxation based on the LDA energy.

The conclusion that the AK13 energetics cannot primarily be attributed to misfeatures in its self-consistent density opens for a hypothetical future functional that mostly retains the AK13 orbitals while improving the energetics. We take a first step towards such a functional by proving that the use of LDA xc on AK13 densities is equivalent to a single consistent xc functional that has an explicit dependence on the external potential. Future work is necessary to turn this approach into a functional fully within KS DFT. However, even without such a functional, our results suggest that LDA on self-consistent AK13 densities works in practice as a computational scheme that successfully combines improved band structure features with predictive lattice constants and bulk moduli. However, the tendency to over-bind is somewhat stronger even than ordinary self-consistent LDA. 

\begin{acknowledgments} 
We thank Stephan K\"ummel and Thilo Aschebrock for insightful discussions, and valuable input on an early version of the manuscript. R.~A. gratefully acknowledges support from the Swedish Research Council (VR), Grant No. 621-2011-4249 as well as the Linnaeus Environment at Link\"oping on Nanoscale Functional Materials (LiLi-NFM) funded by VR.  

\end{acknowledgments} 

    \end{document}